# On unitarity of the particle-hole dispersive optical model

**M.L. Gorelik** [1)], **S. Shlomo** [2,3)], **B.A. Tulupov** [4)], **M.H. Urin** [5)]

1) Moscow Economic School, Moscow 123022, Russia

2) Cyclotron Institute, Texas A&M University, College Station, TX 77843, USA

3) Department of Particles and Astrophysics, the Weizmann Institute of Science, Rehovot 76100, Israel

4) Institute for Nuclear Research, RAS, Moscow 117312, Russia

5) National Research Nuclear University "MEPhI", Moscow 115409, Russia

For the recently developed particle-hole dispersive optical model, weak violations of unitarity due to a phenomenological description of the spreading effect are considered. Methods for unitarity restoration are proposed and implemented for the $^{208}$Pb nucleus in the description of the energy-averaged isoscalar monopole double transition density and strength functions in a wide excitation energy interval that includes the isoscalar giant monopole resonance and its overtone. To illustrate abilities of the model, direct neutron decay of the mentioned giant resonance is also considered.

*Keywords:* Continuum random phase approximation; Isoscalar giant monopole resonance; Transition density; Particle-hole dispersive optical model

*PACS: 21.60.Jz, 24.10.Ht, 24.30.Cz*

# 1. INTRODUCTION

The particle-hole (p-h) dispersive optical model (PHDOM) was developed recently to describe simultaneously the main relaxation modes of high-energy (p-h)-type excitations in medium-heavy mass spherical nuclei [1]. These modes include the p-h strength distribution (Landau damping), coupling to the single-particle (s-p) continuum and to many-quasiparticle configurations (the spreading effect). The model, formulated with the use of the Fermi-system Green functions [2], is, actually, an extension of the standard [3] and non-standard [1, 4] continuum-RPA (cRPA) versions to the description (phenomenological and in average over the energy) of the spreading effect. This latter is considered in terms of the specific p-h interaction (polarization operator, or p-h self-energy term) responsible for the spreading effect. After energy averaging, the strength of this interaction is properly parameterized to satisfy a statistical assumption concerned with independent spreading of different p-h states. This allows one to get in a closed form the expression for the PHDOM basic quantity – the "free" p-h Green function (p-h response function, or p-h propagator). Being the direct extension of the discrete-RPA p-h propagator, this expression contains the imaginary and real parts of the strength of the energy-averaged p-h self-energy term. The imaginary part determines the real one via the proper dispersive relationship, which follows from the spectral expansion of the 2p-2h Green function (2p-2h configurations are the doorway-states for the spreading effect) [1, 5].The "free" p-h propagator corresponds to the model of non-interacting independently damping quasiparticles. Within the PHDOM, the s-p continuum is taken into account with the use of an approximate spectral expansion for the Green function of the Schrodinger equation involving the imaginary and (dispersive) real additions to the mean field. The imaginary part of the combined s-p potential is relatively small (as compared with the imaginary part of the potential used for the optical-model description of nucleon-nucleus scattering) due to a (destructive) interference in spreading of particles and holes. This point has been noted a long ago (see, e. g., Ref. [6]).

The unique feature of the PHDOM is its ability to describe properties of (p-h)-type excitations at arbitrary (but high-enough) energies. These properties are: the p-h strength functions (including distant "tails" of giant resonances); the p-h double transition density – the key quantity in description of hadron-nucleus scattering accompanied by excitation of (p-h)-type states; the partial branching ratios for direct nucleon decay of mentioned states that is accompanied by population of single-hole states of the product nucleus. The first extensive implementations of the PHDOM are concerned with the simplest photo-nuclear reactions [7], and high-energy isoscalar monopole (ISM) excitations [8].

The methods used within the PHDOM for the description of the spreading effect lead to weak violations of the model unitarity. The sources of unitarity violations are an energy dependence of the energy-averaged p-h self-energy term, and also the above-mentioned approximate spectral expansion of the optical-model Green function. Unitarity violation within the s-p optical model due to the mean-field energy dependence is discussed in Refs. [9-11]. The signatures of unitarity violations within the PHDOM are the appearance of: (i) nonzero values (markedly larger than uncertainties of numerical calculations) of the spurious strength function, corresponding to the unit external field; (ii) small negative values of the strength function of the isoscalar giant monopole resonance (ISGMR) at the energies much larger than the ISGMR energy[12]. The last effect leads to an underestimation of the total ISGMR strength.

In this work we investigate weak unitarity violations within the particle-hole dispersive optical model and propose methods for unitarity restoration. The methods are illustrated by consideration of high-energy isoscalar monopole excitations in the $^{208}$Pb nucleus. In particular, we study the energy-averaged isoscalar monopole double transition density and strength functions in a wide excitation-energy interval that includes the isoscalar giant monopole resonance and its overtone. Preliminary results of this investigation are given in Refs. [12, 13]. As an adjunct to studies of Refs. [8, 12, 13], we also consider direct neutron decay of the ISGMR.

## II. METHODS FOR UNITARITY RESTORATION WHITHIN PHDOM

### A. Discrete version of the model ("λ-representation")

The basic PHDOM quantity, the "free" p-h propagator $A_0(x, x',\omega)$, can be expanded in terms of s-p wave functions $\varphi_\lambda(x)$ ("λ-representation" in terminology of Ref. [2]). Here, x is the set of s-p coordinates including spin and isospin variables, ω is the excitation energy. A s-p Hamiltonian $H_0(x)$ determines the s-p energies and wave functions: $(H_0(x) - \varepsilon_\lambda)\varphi_\lambda(x) = 0$. The above-mentioned expansion

$$A_0(x, x', \omega) = \sum_{\lambda\mu} \phi_\mu^*(x)\phi_\lambda(x)\phi_\lambda^*(x')\phi_\mu(x') A_{\lambda\mu}(\omega) \tag{1}$$

contains the expansion elements $A_{\lambda\mu}(\omega)$, which can be represented as the sum of direct (resonance) and backward (non-resonance) terms:

$$A_{\lambda\mu}(\omega) = -\frac{(1-n_\lambda)n_\mu}{\varepsilon_\lambda-\varepsilon_\mu-\omega-a_{\lambda\mu}(\omega)} + \frac{(1-n_\mu)n_\lambda}{\varepsilon_\lambda-\varepsilon_\mu-\omega+a_{\lambda\mu}(\omega)} = A_{\lambda\mu}^r(\omega) + A_{\lambda\mu}^{nr}(\omega) \ . \tag{2}$$

Here, $n_{\lambda,\mu}$ and $\varepsilon_{\lambda,\mu}$ are, respectively, the occupation numbers and energies of s-p states; $a_{\lambda\mu}(\omega) = [iW(\omega) - P(\omega)]f_\lambda f_\mu$, where the real and imaginary parts of the strength of the energy-averaged p-h self-energy term, $P(\omega)$ and $W(\omega)$, are related by a proper dispersive relationship [1, 5]; $f_{\lambda,\mu}$ are the diagonal matrix elements of the Woods-Saxon function, $f(r)$. Since $W(\omega)$ and $P(\omega)$ are even functions of $\omega$, the expansion coefficients of Eq. (2) satisfy the symmetry condition: $A_{\lambda\mu}(\omega) = A_{\mu\lambda}(-\omega)$ .

In the absence of the spreading effect, i.e. within the discrete-RPA ($a_{\lambda\mu}(\omega) \to i0^+$, $A_{\lambda\mu} \to A_{\lambda\mu}^{(0)}$), the corresponding resonance term in Eq. (2) determines the p-h transition-density element $\rho_{\lambda\mu}^{(0)}(\omega) = -\frac{1}{\pi}\text{Im}A_{\lambda\mu}^{(0)}(\omega)$ and the unit p-h strength $\rho_{\lambda\mu}^{(0)} = \int \rho_{\lambda\mu}^{(0)}(\omega)d\omega = (1 - n_\lambda)n_\mu$. In the presence of the spreading effect, the resonance term in Eq. (2) is mainly responsible (within the accuracy of order $(W(\omega)/\omega)^2$) for unitarity violation caused by the energy dependence of the strength of the averaged p-h self-energy term. This weak violation can be approximately described in terms of $dP(\omega)/d\omega$ under the realistic assumption that in average $(dW/d\omega)^2 \ll |dP/d\omega|$ . Under the above assumptions, the transition-density element $\rho_{\lambda\mu}(\omega) = -\frac{1}{\pi}\text{Im } A_{\lambda\mu}(\omega)$ is mainly determined by the resonance term in Eq. (2). This element can be described by a Lorentzian that has a maximum at the energy $\omega_{\lambda\mu} = \varepsilon_\lambda - \varepsilon_\mu + P(\omega_{\lambda\mu})f_\lambda f_\mu$ and whose width and p-h strength are:

$$\Gamma_{\lambda\mu} = 2W(\omega_{\lambda\mu})/p_{\lambda\mu}(\omega_{\lambda\mu}) \ , \qquad \rho_{\lambda\mu} = \rho_{\lambda\mu}^{(0)}/p_{\lambda\mu}(\omega_{\lambda\mu}) \ , \tag{3a}$$

with $p_{\lambda\mu}$ being the diagonal matrix element of

$$p_\lambda(r, \omega) = 1 - dP/d\omega\, f_\lambda f(r) \qquad . \qquad (3b)$$

The expression for $\rho_{\lambda\mu}$ specifies a degree of unitarity violation and suggests that unitarity can be restored by replacing the "free" response-function element $A_{\lambda\mu}(\omega)$ of Eq. (2) by the modified quantity $A^m_{\lambda\mu}(\omega)$:

$$A^m_{\lambda\mu}(\omega) = A^r_{\lambda\mu}(\omega)p_{\lambda\mu}(\omega) + A^{nr}_{\lambda\mu}(\omega)p_{\lambda\mu}(-\omega) \qquad . \qquad (4)$$

As a result, the modified p-h strength is restored, that is $\rho^m_{\lambda\mu} = \rho^{(0)}_{\lambda\mu}$, and $A^m_{\lambda\mu}(\omega) = A^m_{\mu\lambda}(-\omega)$.

It is worth to note that the modified "free" response function $A^m_0(x, x', \omega)$ satisfies, in just the same way as the function $A_0(x, x', \omega)$, the condition $\int A^m_0(x, x', \omega)dxdx' = 0$. In other words, within the "$\lambda$-representation", the unit external field does not generate (isoscalar monopole) excitations. Another comment concerns with neglecting the $W$ energy dependence. After taking the quantity $dW(\omega)/d\omega$ into account, the p-h strength is modified as follows:

$$\rho_{\lambda\mu} = \rho^{(0)}_{\lambda\mu}/p_{\lambda\mu}(\omega_{\lambda\mu})(1 + \delta^2_{\lambda\mu}) \quad ,$$

$$\delta_{\lambda\mu} = (dW/d\omega)_{\omega=\omega_{\lambda\mu}} f_\lambda f_\mu / p_{\lambda\mu}(\omega_{\lambda\mu}) \qquad . \qquad (5)$$

The question of whether the quantities $\delta^2_{\lambda\mu}$ are small is discussed in Sect. 3.

### B. Continuum version of the model

In just the same way, as it is done in Ref [1], we take into account the s-p continuum in the expression for the modified response function $A^m_0(x, x', \omega)$, making the use of an approximate spectral expansion for the Green function of the Schrodinger equation involving a complex-valued energy-dependent potential. With allowance for Eq. (4), we obtain:

$$A^m_0 = A^{m,(1)}_0 + A^{m,(2)}_0 + A^{m,(3)}_0 \quad ,$$

$$A^{m,(1)}_0 = \sum_\mu n_\mu\, \varphi^*_\mu(x) g^m(x, x', \varepsilon_\mu + \omega) \varphi_\mu(x') \quad ,$$

$$A^{m,(2)}_0 = \sum_\lambda n_\lambda\, \varphi^*_\lambda(x') g^m(x', x, \varepsilon_\lambda - \omega) \varphi_\lambda(x) \quad ,$$

$$A^{m,(3)}_0 = \sum_{\lambda\mu} n_\lambda n_\mu \varphi^*_\mu(x)\varphi_\lambda(x)\varphi^*_\lambda(x')\varphi_\mu(x') \left\{ \frac{p_{\lambda\mu}(\omega)}{\varepsilon_\lambda - \varepsilon_\mu - \omega - a_{\lambda\mu}(\omega)} - \frac{p_{\lambda\mu}(-\omega)}{\varepsilon_\lambda - \varepsilon_\mu - \omega + a_{\lambda\mu}(\omega)} \right\} \qquad . \qquad (6)$$

The modified optical-model Green functions in these expressions, $g^m(x, x', \varepsilon_\nu \pm \omega)$, can be presented in the form, used previously in Refs. [10, 11]:

$$g^m(x, x', \varepsilon_\nu \pm \omega) = p^{1/2}_\nu(x, \pm\omega) g(x, x', \varepsilon_\nu \pm \omega)\, p^{1/2}_\nu(x', \pm\omega) \quad , \qquad (7)$$

where the non-modified optical-model Green functions satisfy the equations:

$$\{H_0(x) - [\varepsilon_\nu \pm \omega + (iW(\omega) - P(\omega))f_\nu f(x)]\}g(x, x', \varepsilon_\nu \pm \omega) = -\delta(x - x') \qquad . \qquad (8)$$

Since these Green functions can be presented as a product of the regular and irregular solutions of the homogeneous equation that related to Eq. (8), the modified regular continuum state wave functions corresponding to the relationship (7) are equal to

$$\psi^{(\pm),m}(x, \varepsilon = \varepsilon_\mu + \omega) = p_\mu^{1/2}(x, \omega)\psi^{(\pm)}(x, \varepsilon = \varepsilon_\mu + \omega) \quad . \tag{9}$$

The non-modified wave functions $\psi^{(+)}(x, \varepsilon > 0)$ have been used within the PHDOM for the description of direct-decay properties of giant resonances [1, 7].

Similarly to the initial PHDOM version, the starting point of the modified version is the Bethe-Goldstone-type equation for the modified energy-averaged p-h Green function:

$$A^m(x, x', \omega) = A_0^m(x, x', \omega) + \int A_0^m(x, x'', \omega) F(x'') A^m(x'', x', \omega) \, dx'' \quad . \tag{10}$$

Here, the "free" p-h propagator is determined by Eqs. (6) – (8), $F(x)$ is the strength of the p-h interaction (taken as the Landau-Migdal forces [2]) responsible for long-range correlations. The energy-averaged modified double transition density $\rho^m(x, x', \omega)$ and the strength function $S_{V_0}^m(\omega)$, related to a s-p external field (probing operator) $V_0(x)$, are determined by the p-h Green function of Eq. (10):

$$\rho^m(x, x', \omega) = -\frac{1}{\pi} \operatorname{Im} A^m(x, x', \omega) \tag{11}$$

and

$$S_{V_0}^m(\omega) = \int V_0^+(x)\rho^m(x, x', \omega)V_0(x')dxdx' =$$

$$= -\frac{1}{\pi} \operatorname{Im} \int V_0^+(x)A_0^m(x, x', \omega)V^m(x', \omega)dxdx' \quad . \tag{12}$$

Here, $V_0^+(x)$ is the hermitien conjugated operator and the modified effective field $V^m(x, \omega)$ satisfies the equation:

$$V^m(x, \omega) = V_0(x) + F(x) \int A_0^m(x, x', \omega) V^m(x', \omega) \, dx' \quad . \tag{13}$$

The modified decay-channel strength function $S_{V_0,c}^m(\omega)$ ($c$ is the set of decay-channel quantum numbers) is determined by the modified effective field of Eq. (13) as follows:

$$S_{V_0,c}^m(\omega) = n_\mu | \int \psi^{(-),m*}(x, \varepsilon = \varepsilon_\mu + \omega) V^m(x, \omega) \psi_\mu(x) dx$$

$$\int \psi_\mu^*(x') V^{m*}(x', \omega) \psi^{(+),m}(x', \varepsilon = \varepsilon_\mu + \omega) dx' | \quad , \tag{14}$$

where the modified continuum-state wave functions are determined by Eq. (9).

In the limit $p_v(r, \omega) \to 1$, Eqs. (6) – (14) are naturally related to the initial continuum-PHDOM version of Ref. [1]. It is worth to note here, that within this version, the continuum-state wave functions $\psi^{(+)}(x, \varepsilon)$ are used instead of $\psi^{(-)*}(x, \varepsilon)$ in expressions related to the output channel of reactions with one nucleon in the continuum. Such a substitution is valid only within the cRPA.

The equations of this Subsection are applicable (after separation of spin-angular and isospin variables) for the description of high-energy (p-h)-type excitations having arbitrary values of angular momentum, parity, isospin and its third projection. Only isoscalar monopole excitations have need for a special consideration, which is done in the next Subsection.

### C. PHDOM unitary version for describing isoscalar monopole excitations

As mentioned in the Introduction, the use within the model of an approximate spectral expansion for the optical-model Green functions (see the resulted Eqs. (7),(8)) leads to appearance of "spurious" (ISM) excitations generated by the unit external field. The removal of corresponding unitarity violation is meaningful in describing only of ISM excitations, but this is of particular interest since the nuclear matter incompressibility coefficient depends on the energy centroid of the ISGMR.

Within the PHDOM, properties of ISM excitations are determined by the corresponding radial component of the double transition density $\rho(r, r', \omega)$ [8]. In particular, the strength function $S_k(\omega)$ related to an arbitrary ISM external field $V_{0,k}(\vec{r}) = V_{0,k}(r)Y_{00}$ equals to

$$S_k(\omega) = \int V_{0,k}(r)\rho(r,r';\omega)V_{0,k}(r')drdr' \quad . \tag{15}$$

Hereafter, we consider the strength functions for the ISGMR ($V_{0,1}(r) = r^2$), its overtone, ISGMR2 ($V_{0,2}(r) = r^4 - \eta r^2$, parameter $\eta$ is defined in Ref. [8]), and the "spurious" strength function ($V_{0,0}(r) = 1$).

To remove unitarity violation caused by appearance of nonzero values of $S_0(\omega)$, we use the renormalized ISM double transition density $\rho^{rn}(r, r', \omega)$, which is defined by the condition $S_0^{rn}(\omega) = 0$. This condition is fulfilled, provided that the mentioned density is taken in the form:

$$\rho^{rn}(r,r',\omega) = \rho(r,r',\omega) - \bar{n}(r) \int \rho(r,r',\omega)dr - \\ -\bar{n}(r') \int \rho(r,r',\omega)dr' + \bar{n}(r)\bar{n}(r')S_0(\omega) \quad . \tag{16}$$

Here, $\bar{n}(r) = 4\pi r^2 n(r)/A$ is the radial nuclear matter density for the nucleus, containing $A$ nucleons. Being normalized to unity, $\int \bar{n}(r)dr = 1$, this density can be considered as the radial transition density of the ISM "spurious" excitation.

The renormalized strength function, $S_k^{rn}(\omega)$, corresponding to the radial external field $V_{0,k}(r)$ is an important consequence of the transition to the renormalized ISM double transition density of Eq. (16). In accordance with Eqs. (15), (16), we get

$$S_k^{rn} = \int V_{0,k}(r)\rho^{rn}(r,r',\omega)V_{0,k}(r')drdr' = \int \delta V_{0,k}(r)\rho(r,r',\omega)\delta V_{0,k}(r')drdr' \quad , \tag{17}$$

where $\delta V_{0,k}(r) = V_{0,k}(r) - \bar{V}_{0,k}$ with $\bar{V}_{0,k} = \int V_{0,k}(r)\bar{n}(r)dr$. Thus, the renormalization procedure, required for calculating $S_k^{rn}(\omega)$ on the base of Eq. (15), can be performed in two equivalent ways: the first one consists in replacing $\rho(r, r', \omega)$ by $\rho^{rn}(r, r', \omega)$ in this equation, another – in replacing $V_{0,k}(r)$ by $\delta V_{0,k}(r)$. The second way is similar to the procedure of eliminating the "spurious"-state contribution to observables. An appropriate example here might be elimination of the contribution to the strength function of the isoscalar dipole giant resonance from the $1^-$ "spurious" state corresponding to center-of-mass motion (see, e.g., Ref. [14]).

The alternative and more simple in practical realization method for calculating the strength function (17) is based on the use of the effective field that satisfies the equation

$$\delta V_k(r,\omega) = \delta V_{0,k}(r) + \frac{F(r)}{r^2} \int A_0(r,r',\omega) \delta V_k(r',\omega) dr' \tag{18}$$

($F(r)$ is the radial-dependent strength of the isoscalar part of the Landau-Migdal p-h interaction), and determines the strength function accordingly:

$$S_k^{rn}(\omega) = -\frac{1}{\pi} \operatorname{Im} \int \delta V_{0,k}(r) A_0(r,r',\omega) \delta V_k(r',\omega) dr dr' \quad . \tag{19}$$

Here, $A_0$ is the radial component of the (nonmodified) "free" p-h propagator. The effective field of Eq. (18) determines the partial decay-channel strength function as follows:

$$S_{k,\mu}^{rn}(\omega) = (2j_\mu+1) n_\mu \ |\int \chi_{(\mu)}^{(-),*}(r, \varepsilon_\mu + \omega) \delta V_k(r,\omega) \chi_\mu(r) dr$$

$$\int \chi_\mu(r') \delta V_k^*(r',\omega) \chi_{(\mu)}^{(+)}(r', \varepsilon_\mu + \omega) dr'| \quad . \tag{20}$$

Here, $r^{-1}\chi_\mu(r)$ is the radial bound-state s-p wave function and $r^{-1}\chi_{(\mu)}^{(\pm)}(r,\varepsilon)$ are the radial continuum-state s-p wave functions $((\mu) = j_\mu, l_\mu)$. The above-given Eqs. (18) – (20) represent the specific case of Eqs. (12) – (14).

The strength function of Eq. (20) can be used to evaluate the partial branching ratio for direct nucleon decay of the ISGMR and its overtone accompanied by population of the single-hole state $\mu^{-1}$ of the product nucleus:

$$b_{k,\mu}^{rn}(\delta) = \int_{(\delta)} S_{k,\mu}^{rn}(\omega) d\omega / \int_{(\delta)} S_k^{rn}(\omega) d\omega \quad . \tag{21}$$

Here, integration is performed over an excitation-energy interval δ that includes the given giant resonance.

Together with the renormalization of the above-considered ISM quantities, it is necessary to take into account modification of these quantities caused by the energy dependence of the strength of the p-h self-energy term (Subsect. II.B). The expressions for the corresponding doubly-modified quantities – double transition density $\rho^u(r, r', \omega)$, strength functions $S_k^u(\omega)$, partial decay-channel strength functions $S_{k,\mu}^u(\omega)$ – can be obtained by proper substitutions in Eqs. (16)-(20). They are: $\rho(r, r', \omega) \to \rho^m(r,r',\omega)$ in Eq. (16) to get $\rho^u(r,r',\omega)$; $\rho^{rn}(r,r',\omega) \to \rho^u(r,r',\omega)$ in Eq. (17) to get $S_k^u(\omega)$; $A_0(r,r',\omega) \to A_0^m(r,r',\omega)$ in Eqs. (18) and (19) to get $\delta V_k^m(r,\omega)$ and the alternative representation for $S_k^u(\omega)$; $\chi_{(\mu)}^{(\pm)}(r, \varepsilon = \varepsilon_\mu + \omega) \to \chi_{(\mu)}^{(\pm),m}(r, \varepsilon = \varepsilon_\mu + \omega)$ and $\delta V_k(r,\omega) \to \delta V_k^m(r,\omega)$ in Eq. (20) to get $S_{k,\mu}^u(\omega)$. The expression for the partial branching ratio $b_{k,\mu}^u$ can be properly obtained from Eq. (21). These statements complete the transition to the unitary version of the model (abbreviated below as PHDOM-UV) in applying to the description of ISM excitations.

## III. DESCRIPTION OF ISOSCALAR MONOPOLE EXCITATIONS IN THE $^{208}$PB PARENT NUCLEUS WITHIN THE PHDOM UNITARY VERSION

The PHDOM initial version has been implemented in Ref. [8] to describe high-energy ISM excitations in the $^{208}$Pb nucleus. In particular, the energy-averaged radial component of the ISM double transition density $\rho(r, r', \omega)$ has been analyzed in a wide excitation-energy interval that includes the ISGMR and its overtone ISGMR2. Other subjects for study have been the relative energy-weighted strength functions

$$y_k(\omega) = \omega S_k(\omega)/(EWSR)_k \ , \tag{22}$$

where $S_k(\omega)$ and $(EWSR)_k$ are, respectively, the strength functions of Eq. (15) for the mentioned giant resonances and the sum rules related to the corresponding probing operators $V_{0,k}(r)$. Methods for calculating the above quantities within the PHDOM initial version and the choice of model parameters are described in details in Ref. [8]. In this reference, the relaive strength functions $y_k(\omega)$ have been described in a rather narrow interval around the maximum of the respective giant resonance. However, in considering large intervals, there appear rather small (in absolute values) negative (unphysical) values of the strength functions $y_k(\omega)$. As applied to the ISGMR, this statement is illustrated in Fig. 1 (see also Ref. [12]). The negative values of $y_1(\omega)$ lead to an underestimation of the total ISGMR strength. Hereafter, the shown calculated results are obtained using the set of model parameters exploited in Ref. [8].

Since the method used within the PHDOM initial version for describing the spreading effect leads to violations of model unitarity, before presentation of calculation results we show the used parameterization of the phenomenological quantity $W(\omega)$ and the respective (dispersive) quantity $P(\omega)$ [5,1]:

$$2W(\omega) = \begin{cases} 0 & (\omega \leq \Delta) \\ \alpha \frac{(\omega-\Delta)^2}{1+(\omega-\Delta)^2/B^2} & (\omega \geq \Delta) \end{cases} \tag{23}$$

and

$$P(\omega) = \frac{2}{\pi} \text{ P.V.} \int_\Delta^\infty W(\omega') \left( \frac{\omega'}{\omega^2-\omega'^2} - \frac{\omega'}{\Delta^2-\omega'^2} \right) d\omega' \ . \tag{24}$$

The explicit expression for $P(\omega)$ is rather combersome and can be found in Ref. [5]. The "gap" parameter $\Delta=3$ MeV and the "saturation" parameter $B=7$ MeV in Eq. (23) are taken universal for medium-heavy mass spherical nuclei (see, e.g., Ref. [7]), while the strength parameter $\alpha = 0.07$ MeV has been adjusted in Ref. [8] to describe within the PHDOM the observable total width of the ISGMR in $^{208}$Pb. The quantities of Eqs. (23) and (24) are shown in Fig. 2. In Figs. 3, 4 and 5, the diagonal double transition density $\rho^u(r, r, \omega)$ (taken for different values of the energy), the relative strength functions $y_k^u(\omega)$ calculated within the PHDOM-UV for the ISGMR and ISGMR2 are given in a comparison with the respective quantities obtained within the PHDOM initial version. These results can be considered as an evidence of weak violation of model unitarity. Nevertheless, restoration of model unitarity allows one, in particular, to eliminate negative values of the ISGMR relative strength function (Fig. 1).

As a measure of unitarity violation, we use the integrated quantities $x_k = \int y_k(\omega)d\omega$ found for the wide energy interval 3-83 MeV. Table 1 gives these quantities calculated within: (i) the cRPA, i. e. in the approximation $W = P = 0$ ($x_k^{cRPA}$); (ii) the PHDOM initial version ($x_k^i$); (iii) the version modified only by taking $dP/d\omega$ into account ($x_k^m$); (iv) the version that employs only the renormalized double transition density ($x_k^{rn}$); (v) the PHDOM-UV ($x_k^u$). From the calculation results obtained by the use of relationships given in Sect 2 and presented in Table 1 and Fig. 1, it follows that, in describing high-energy ISM excitations in the $^{208}$Pb nucleus, one can eliminate weak violations of unitarity within the PHDOM initial version. Also, the contributions from two sources of unitarity violation are partially cancelled. It is noteworthy that the explicit inclusion into consideration of the *W* energy dependence (see Eqs. (5)) gives a relatively small contribution to the unitarity restoration. This statement follows f≈rom a comparison of the averaged values of $dP/d\omega$ and $\delta^2(\omega)$. The calculated values of the quantities $\pi_k = \int \frac{dP(\omega)}{d\omega} y_k^u(\omega) d\omega$ and $w_k = \int \delta^2(\omega) y_k^u(\omega) d\omega$ are given in Table 1. In conclusion of the above-given analysis, we show also the calculated quantities $x_0^i$ and $x_0^u$, which determine the "spurious" strengths: $x_0^i = \int |S_0^i(\omega)| d\omega \sim 0.3 \cdot 10^{-2}$ and $x_0^u = \int |S_0^u(\omega)| d\omega \sim 0$ (within uncertainties of numerical calculations).

The branching ratios for ISGMR direct neutron decay, accompanying by population of the single-hole states $\mu^{-1}$ in the $^{207}$Pb product nucleus, calculated in accordance with the relationships given in Sect. 2 within the PHDOM initial and unitary versions, $b_\mu^i$ and $b_\mu^u$, respectively, are given in Table 2 in a comparison with the respective experimental data of Refs.[16, 17]. The branching ratios for ISGMR direct proton decay are negligibly small. The comment to the calculated quantities is concerned with the exploited calculation scheme, in which pure single-quasiparticle nature of the product-nucleus single-hole states is assumed. Actually, due to particle-phonon coupling, the single-hole strength of low-energy $\mu^{-1}$ states might be somewhat decreased. For this reason, the calculated branching ratios can be considered as an upper limit for the corresponding experimental values. The calculated total branching ratios $b_{tot} = \sum_\mu b_\mu$ are in agreement with qualitative estimation $b_{tot} \approx \Gamma^{cRPA}/\Gamma \approx 0.5$. Here, $\Gamma^{cRPA} \approx 1.5$ MeV is the total width of the ISGMR strength function calculated within the cRPA [8], $\Gamma = 2.88$ MeV is the ISGMR experimental total width used in Ref. [8] to adjust the parameter $\alpha$ in Eq. (23). Thus, the calculated branching ratios are markedly larger than the corresponding experimental values of Refs.[16,17]. Reasons for this discrepancy are not clear now.

## IV. CONCLUDING REMARKS

Fo r the recently developed particle-hole dispersive optical model, we have analyzed unitarity violations, which are due to the phenomenological description of the spreading effect, have formulated methods for unitarity restoration, and have implemented these methods to describing high-energy isoscalar monopole excitations in the $^{208}$Pb nucleus. It has been found that unitarity violations are weak due to a rather small value of the strength of the energy-averaged particle-hole self-energy term responsible for the spreading effect. Nevertheless, such a study seems to be necessary in view of a large number of implementations of the model. We have demonstrated the unique abilities of the mentioned model by considering along with the isoscalar monopole double transition density also the partial branching ratios for direct neutron decay of the isoscalar giant monopole resonance in the $^{208}$Pb.

## ACKNOWLEGEMENTS

This work is supported in part by the Russian Foundation for Basic Research under grant no. 15-02-08007 (M. L. G., B. A. T., M. H. U), by the US Department of Energy under contract # DOE-FG03-93ER40773 (S. S.), and by the Competitiveness Program of NRNU "MEPhI" (M. H. U.). S. S. also thanks the Weizmann Institute of Science for the Weston Visiting Professorship Award and the nice hospitality extended to him.

**References**

[1] M.H. Urin, Phys. Rev. C **87,** 044330(2013).

[2] A.B. Migdal, *Theory of Finite Fermi Systems and Applications to Atomic Nuclei*
    (Interscience, New York, 1967).

[3] S. Shlomo, G. F. Bertch, Nucl. Phys. A **243**, 507 (1975).

[4] M.H. Urin, Nucl. Phys. A **811**, 107 (2008).

[5] B.A. Tulupov, M.H. Urin, Phys. At. Nucl. **72**, 737 (2009).

[6] G.F. Bertch, P.F. Bortignon, R.A. Broglia, Rev. Mod. Phys. **55**, 287 (1983).

[7] B.A. Tulupov, M.H. Urin, Phys. Rev. C **90**, 034613 (2014).

[8] M.L. Gorelik, S. Shlomo, B.A. Tulupov, and M.H. Urin, Nucl. Phys. A **955,** 116 (2016).

[9] C. Mahaux and R. Sartor, Adv. Nucl. Phys. **20**, 1 (1991).

[10] S. Shlomo, V.M. Kolomietz, and Dojbakhsh, Phys. Rev. C **55**,1972 (1997).

[11] G.V. Kolomiytsev, S.Yu. Igashov, M.H. Urin, Phys. At. Nucl. **80, 614** (2017).

[12] M.L. Gorelik, S. Shlomo, B.A. Tulupov, M.H. Urin, EPJ Web Conf. **107**, 05008 (2016).

[13] M.L. Gorelik, B.A. Tulupov, M.H. Urin, Phys. At. Nucl. **79**, 924 (2016).

[14] M.L. Gorelik, S.Shlomo, and M.H.Urin, Phys. Rev. C **62**, 044301 (2000).

[15] M. L. Gorelik and M. H. Urin, Phys. Rev. C **64**, 047301 (2001);
    M. L. Gorelik, M. H. Urin, Phys. At. Nucl. **66**, 1883 (2003).

[16] S. Brandenburg et al., Phys. Rev. C **39**, 2448 (1989).

[17] A. Bracco et al., Phys. Rev. Lett., **60**, 2603 (1988);
    G. Colo et al., Physiscs Letter B **276**, 279 (1992).

**Table 1.** Integrated features of the isoscalar giant monopole resonance ($k=1$) and its overtone ($k=2$) calculated in various approximations

|     | $x_k^{cRPA}$ | $x_k^i$ | $x_k^m$ | $x_k^{rn}$ | $x_k^u$ | $\pi_k$ | $w_k$ |
|-----|--------------|---------|---------|------------|---------|---------|-------|
| k=1 | 0.987        | 0.974   | 0.878   | 1.046      | 1.000   | 0.074   | 0.017 |
| k=2 | 0.984        | 1.047   | 0.999   | 1.062      | 1.001   | 0.049   | 0.005 |

**Table 2.** The partial and total branching ratios (%) for direct neutron decay of the ISGMR in $^{208}$Pb calculated within initial and unitary versions of the PHDOM for the excitation energy interval δ=12.5 – 15.5 MeV, and, respectively, in a comparison with the corresponding quantities obtained from the experimental data of Ref. [16] (a) and Ref. [17] (b).

| $\mu^{-1}$ | $b_\mu^i$ | $b_\mu^u$ | $b_\mu^{\exp(a)}$ | $b_\mu^{\exp(b)}$ |
|---|---|---|---|---|
| $3p_{1/2}$ | 3.2 | 3.6 | 2.9±1.3 | 4.6±1.2 |
| $1i_{13/2}$ | 0.4 | 0.5 | incl.($3p_{1/2}$) | incl.($3p_{1/2}$) |
| $2f_{5/2}$ | 19.5 | 20.6 | 2.6±1.7 | < 2.3±0.5 |
| $3p_{3/2}$ | 6.4 | 7.1 | 5.3±1.8 | 1.6±0.3 |
| $2f_{7/2}$ | 21.5 | 24.7 | 11.4±1.3 | 5.4±1.3 |
| $2d_{5/2}+1g_{7/2}$ | 0.0 | 0.0 | incl.($2f_{7/2}$) | |
| $b_{tot}$ | 51 | 57 | 22 ± 6 | 14 ± 3 |

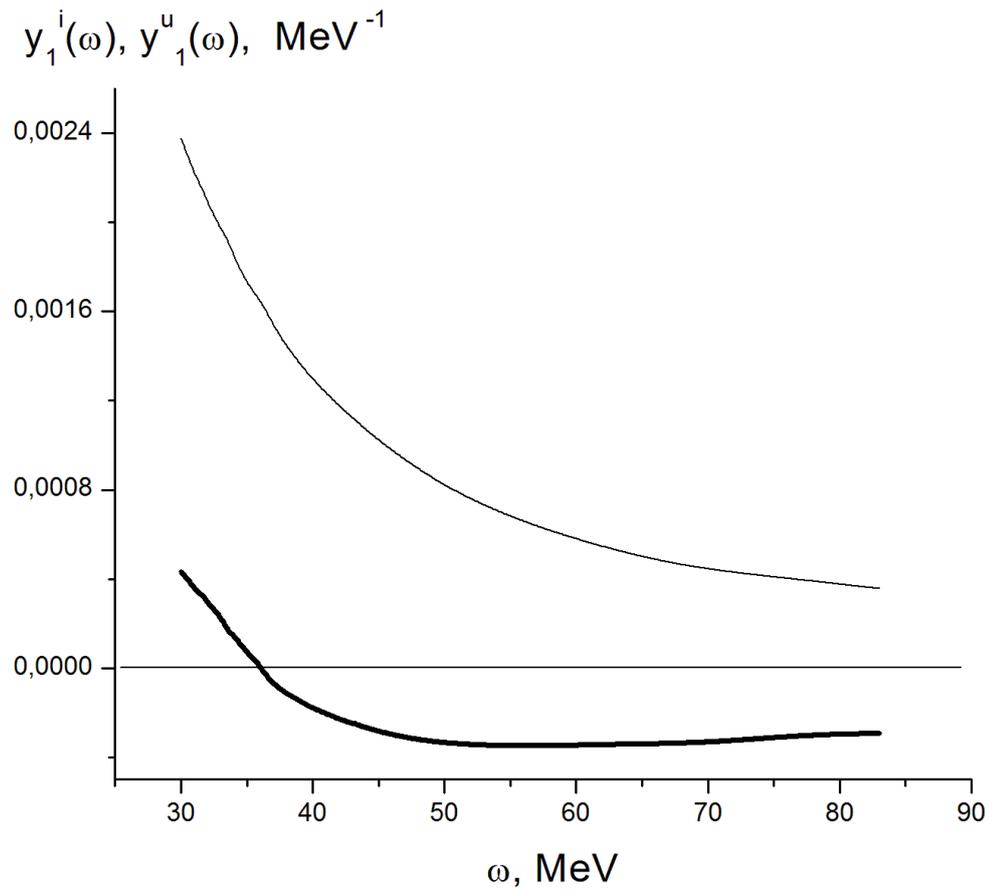

**Fig. 1.** The energy-weighted relative strength functions $y_1^i(\omega)$ (thick line) and $y_1^u(\omega)$ (thin line) calculated for the ISGMR in $^{208}$Pb within, respectively, the initial and unitary versions of the PHDOM.

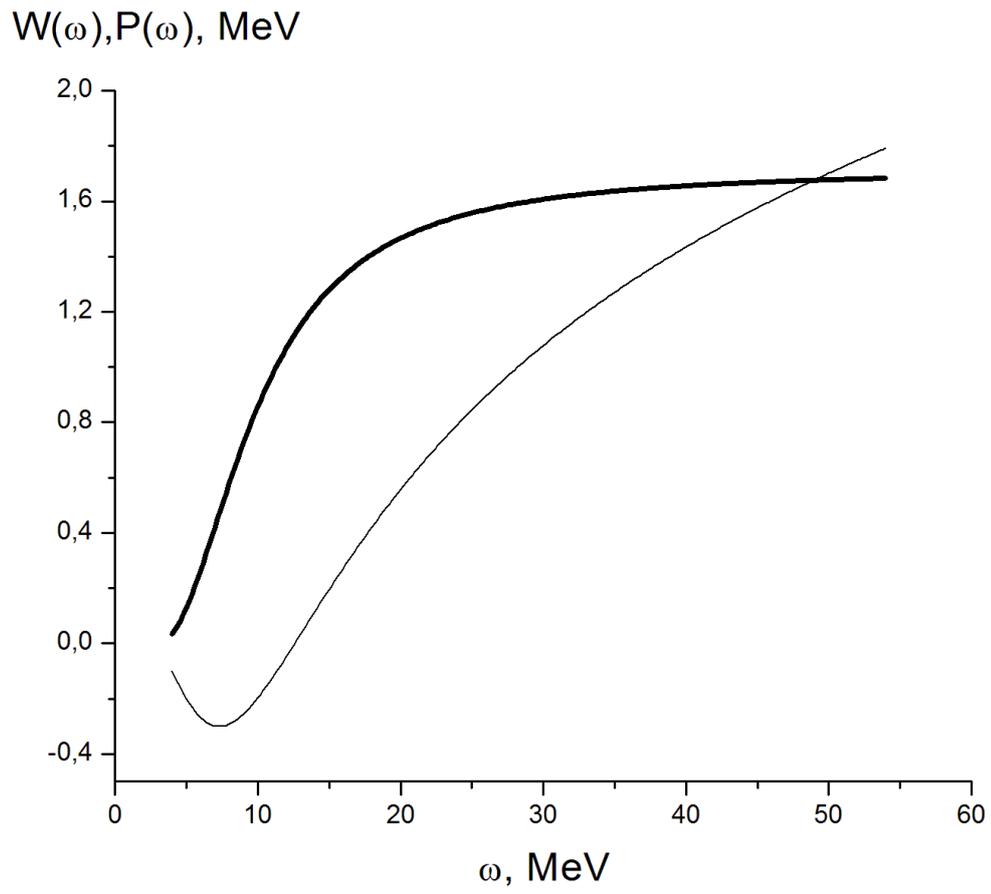

**Fig. 2**. The energy dependence of the imaginary $W(\omega)$ (thick line) and real $P(\omega)$ (thin line) parts of the strength of the energy-averaged p-h self-energy term.

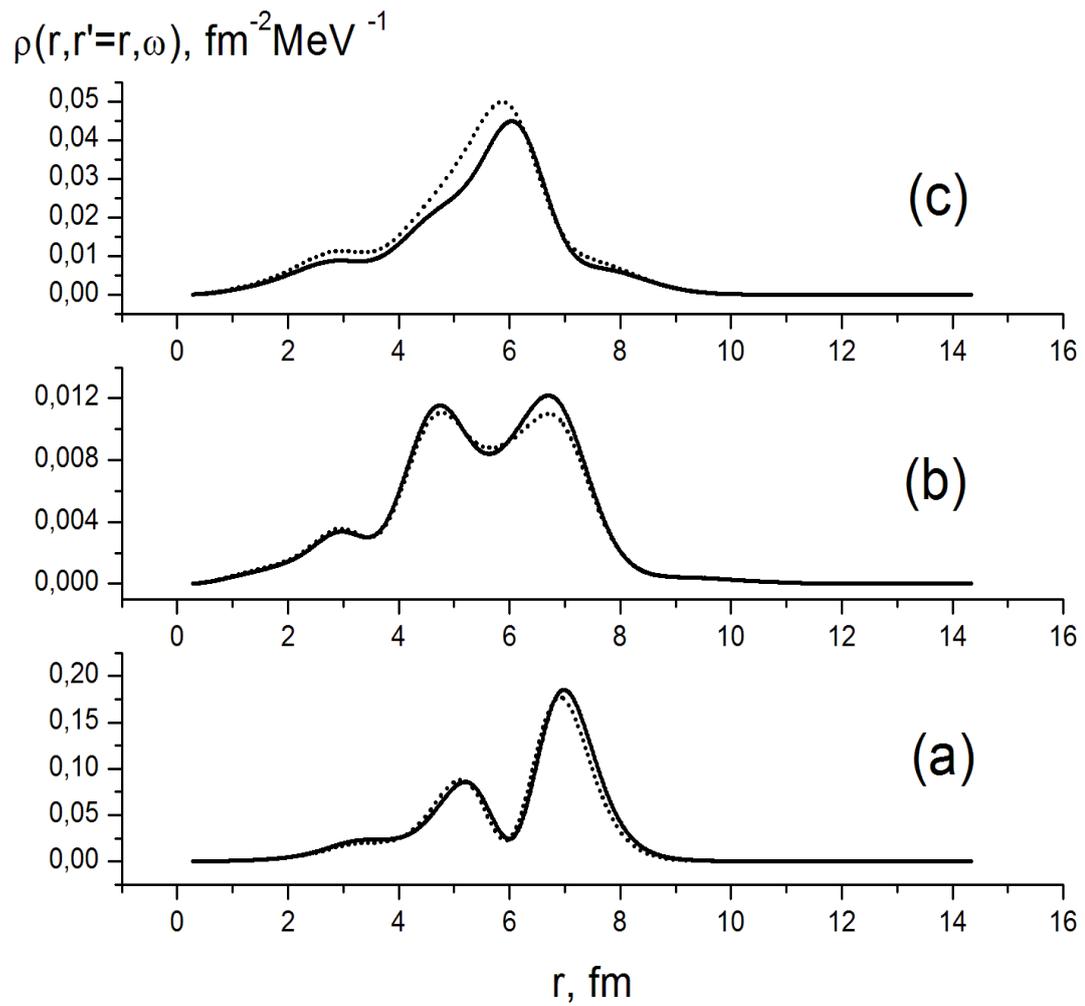

**Fig. 3.** The diagonal energy-averaged radial component of the ISM double transition density calculated for $^{208}$Pb at the different energies : 13.8 MeV (a), 23 MeV (b) and 33 MeV (c) within the initial (thick line) and unitary (dotted line) versions of the PHDOM.

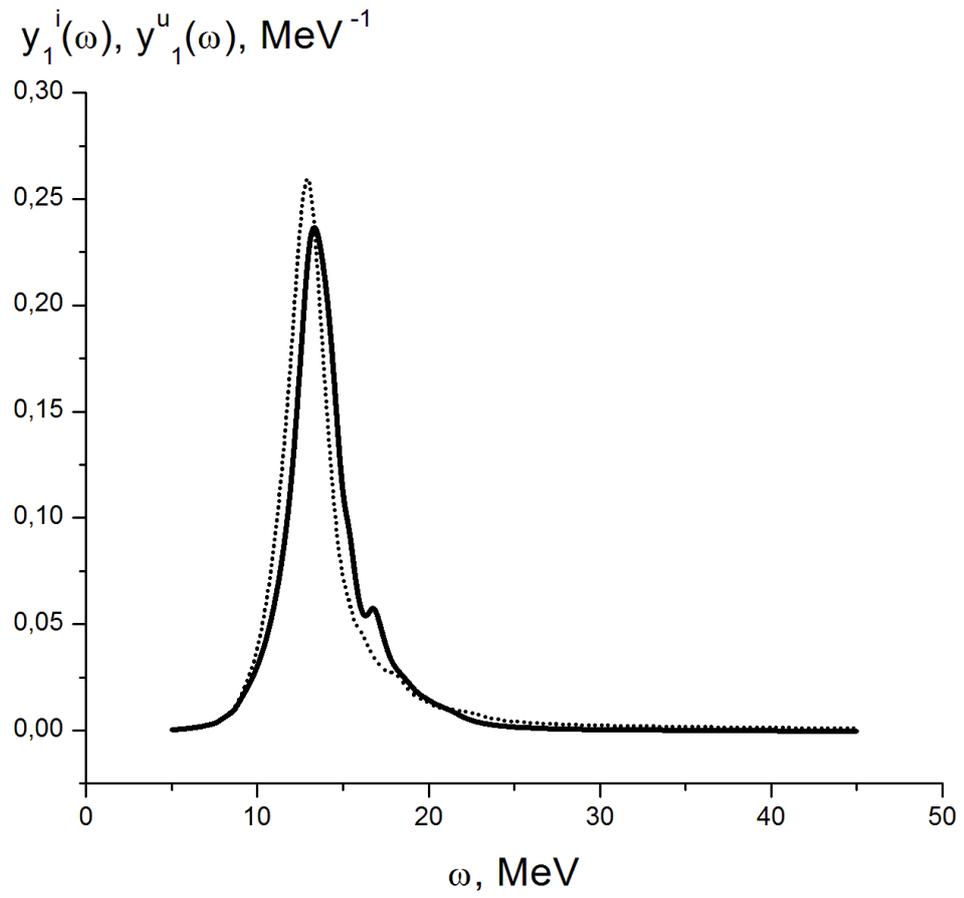

**Fig. 4.** The relative energy-weighted strength functions calculated for the ISGMR in $^{208}$Pb within the initial (thick line) and unitary (dotted line) versions of the PHDOM.

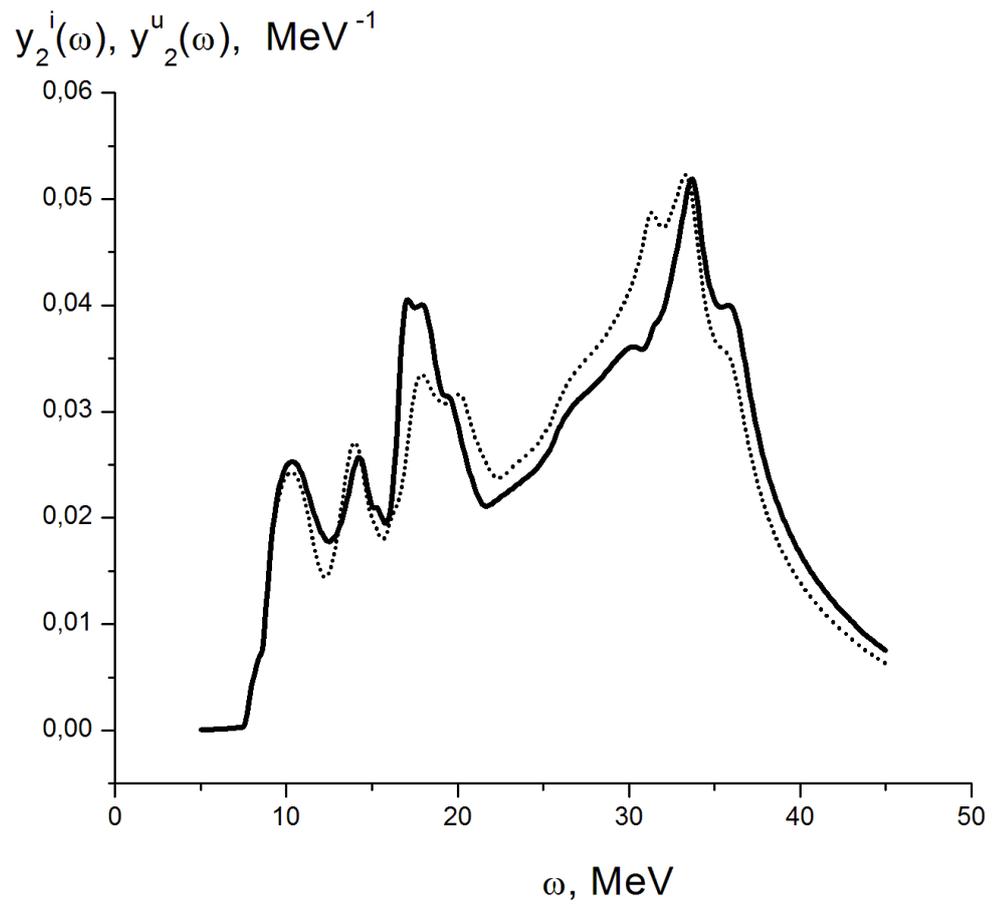

**Fig. 5.** Same as in Fig. 4 but for the ISGMR2.